\documentclass{appolb}
\usepackage{graphicx}
\usepackage{lscape}
\usepackage{amssymb}
\usepackage{amsfonts}
\usepackage{mathrsfs}
\usepackage[numbers,sort&compress]{natbib}

\renewcommand{\thetable}{\Roman{table}}

\begin{document}

\title{Low-$|t|$ structures in elastic scattering at the LHC
}
\author{L\'aszl\'o Jenkovszky\footnote{jenk@bitp.kiev.ua}\\[-0.3cm]
\address{ Bogolyubov Institute for Theoretical Physics, National Academy of Sciences of Ukraine, Kiev, 03680 Ukraine}\\[0.4cm]
{Alexander Lengyel}\footnote{alexander-lengyel@rambler.ru}\\[-0.3cm]
\address{Institute of Electron Physics, National Academy of Sciences of Ukraine, Uzhgorod, 88017 Ukraine}
}
\maketitle
\begin{abstract}
Possible low-$|t|$ structures in the differential cross section of $pp$ elastic scattering at the LHC  are  predicted. It is argued that the change of the slope of the elastic cross section near $t=-0.1$ GeV$^2$ has the same origin as that observed in 1972 at the ISR, both related to the $4m_{\pi}^2$ branch point in the $|t|$-channel of the scattering amplitude. Apart from that structure, tiny oscillations at small $|t|$ may be present on the cone at low $|t|$.
\end{abstract}
\PACS{11.55.Jy, 13.85.Dz, 12.40.Nn}
  
\section{Introduction}
Followed by the first publications on $pp$ elastic scattering at $\sqrt s=7$ TeV
in the broad $\left|t\right|$ range $ 5\cdot10^{-3}\,$GeV$^2<|t|< 2.5\,$GeV$^2$ \cite{TOTEM2,TOTEM1}, 
the TOTEM Collaboration recently made public \cite{Mirko,Simone,Kaspar,Csorgo} their new results at still lower values of $|t|$
at $\sqrt s=8$ TeV. 


Contrary to earlier statements \cite{TOTEM1}, considerable deviation from the linear exponential cone was found. Namely, a change of the 
local slope $B(t)=\frac{d}{dt}($ln$ \frac{d\sigma(s,t)}{dt})$ 
at $8$ TeV by about $0.5$ GeV$^{-2}$ around $|t|\approx 0.1$ GeV$^2$ was observed. 

In the present paper 
we argue that this structure is a recurrence of the similar phenomenon observed in 1972 at the ISR, both related to $t$-channel unitarity effects of the scattering amplitude. Anticipating the relevant TOTEM publication, below we present a method of handling possible structures in the diffraction cone and make predictions based on an a Regge-pole model-based extrapolation from the ISR energy region to that of the LHC. \footnote {Preliminary results of this paper were presented in June, 2014 at the Protvino Conference on High-Energy Physics \cite{Protvino}.}

 \section{The "break" phenomenon; preliminaries}

\begin{figure}[ht]
\center{
\includegraphics[width=.4\textwidth]{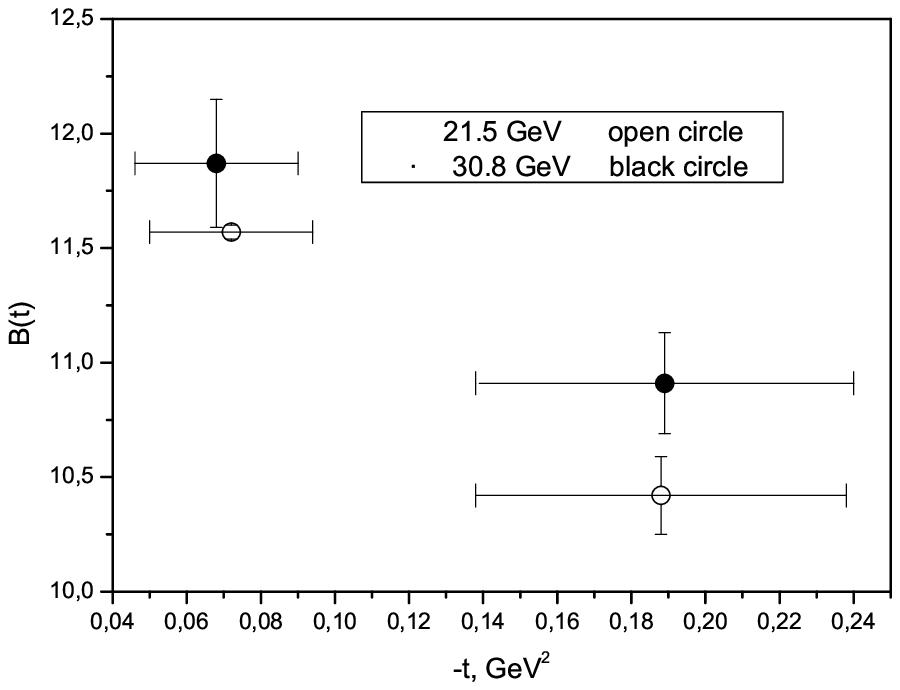}
\includegraphics[width=.4\textwidth]{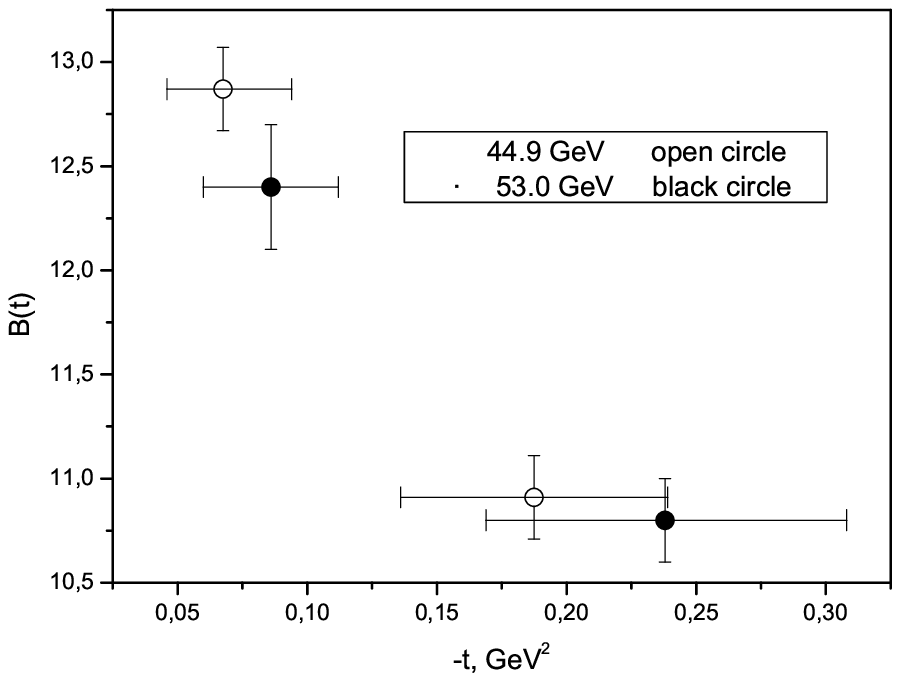}
\caption{ Local slopes $B(t)$ calculated at ISR energies with two different exponentials (see Table~\ref{TabI}).}
}
\label{fig:ISRslope}
\end{figure}

The change of the local slope $B(t)$ around $|t|\approx 0.1$ by about $\approx2\,$GeV$^{-2}$, called the "break" (in fact, a smooth curvature in $B(t)$), at $\sqrt{s}=21.5,$ $23.9,$ $30.8,$  $44.9,$ $53.0\,$GeV) was first observed and discussed in Ref.~\cite{BB72}  (see Table~\ref{TabI} in \cite{BB72}, quoted below and illustrated in Fig.\ref{fig:ISRslope}).

\begin{table}[!hb]
 \centering
 \caption{Slope  B(t) calculated  with different exponential for the ISR data \cite{BB72}.}
 \label{TabI}
\begin{tabular}{|c|c|c|c|c|c|}
  \hline
  E c.m.,  & $\left|t\right|$- range,  & B,  & err. &  $\Delta$ B,  & err.  \\
   GeV & $(\mathrm{GeV}^2)$  &  $(\mathrm{GeV}^{-2})$ &  &  $ (\mathrm{GeV}^{-2})$ &   \\
  \hline
    21.5      & 0.05 - 0.094   & 11.57 &   0.030  &       &               \\
              & 0.138 - 0.2380 & 10.42 &  0.17    & 1.15  &   0.20      \\     
   \hline    
      30.8    & 0.046 - 0.090  & 11.87  &  0.28   &       &              \\
              & 0.138 - 0.240  & 10.91  &  0.22   & 0.96  &   0.50      \\
   \hline
      44.9    & 0.046 - 0.089  & 12.87 &  0.20    &        &                 \\
              & 0.136 - 0.239  & 10.83 &  0.20    &  2.04  &   0.40         \\
   \hline
    53.0      & 0.060 - 0.112  & 12.40 &  0.30    &        &                       \\
              & 0.168 - 0.308  & 10.80 &  0.20    &  1.6   &   0.50     \\
   \hline 
   \end{tabular} 
\end{table}

\begin{figure}[ht]
\center{\includegraphics[width=.4\textwidth]{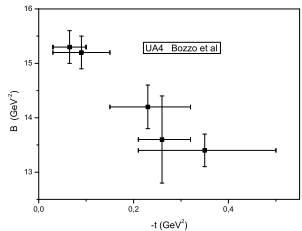}}
\caption{Local slopes B(t) calculated for the UA4 data \cite{BozzoUA4}} 
\label{fig:BozzoUA4}
\end{figure}

 This phenomenon is visible at other energies, see Table~\ref{TabII} and Fig.~\ref{fig:BozzoUA4} {\cite{BozzoUA4}}.
 
\begin{table}
\centering
\caption{Average slope values for fits in different bins of $\left|t\right|$ for the UA4 data \cite {BozzoUA4}.}
 \label{TabII}
\begin{tabular}{|c|c|c|c|}
  \hline
   & $\left|t\right|-$ range,  $(\mathrm{GeV}^2)$ & B,  $(\mathrm{GeV}^{-2}$)  &  err.$(\mathrm{GeV}^{-2})$  \\
  \hline
    I& 0.03 - 0.10  & 15.3 &  0.3  \\
    \hline
   II& 0.03 - 0.15  & 15.2 &  0.2  \\     
   \hline    
  III& 0.15 - 0.32  & 14.2 &  0.4  \\
    \hline
   IV& 0.21 - 0.32  & 13.6 &  0.8  \\
   \hline
   V&  0.21 - 0.50  & 13.4 &  0.3  \\          
   \hline 
   \end{tabular}
\end{table}
   
The magnitude of the slope break at different  ISR energies, calculated in a simple exponential approximation for nearly the same range of momentum transfer $0.05$ GeV$^2)<\left|t\right|< 0.10 (GeV^2)$ and  
  $0.14$ GeV$^2<\left|t\right| < 0.25 GeV^2$ varies within the range $\Delta B(t)\approx 1 - 2 $ GeV$^2$. Using the data from Table~\ref{TabII} on the slope parameter for the SPS energy $546$ GeV for antiproton-proton scattering, one can obtain different values for the slope break depending on the choice of bin pairs. For example, the largest value of the slope break $\Delta B = 1.9 \pm 0.6$ can be obtained by choosing a couple (adjacent bins) made of the farthest values I, V, and the smallest one for the pairs I and III,  $\Delta B = 1.1 \pm 0.7$, taking into account two adjacent intervals.
A different approach to the choice of the individual t-bins  was used in the recent paper \cite{TOTEM1}, where constancy of the slope was stated, at least until 
  $\left|t\right| = 0.2$  GeV$^2$.  
  
 Note that the break found at 8 TeV can be obtained even with simpler methods  
  where two single exponential are fitted in non-overlapping t-ranges, the relevant $B(t)$ differing by more than 7$\sigma$. 
  The overall behaviour of B(t) as a function of  energy is illustrated in \cite{Desgrolard} and \cite{JLL}.

The `break' phenomenon has a clear physical interpretation: it results from the $t-$ channel branch point at $4m_{\pi}^2\approx 0.08$ GeV$^2$ imposed by unitarity. The "break" due to the two pion threshold is related to the pionic atmosphere (cloud) of the nucleon \cite{Cohen} (for more details see next Section).

\begin{figure}[ht]
\center{\includegraphics[width=.6\textwidth]{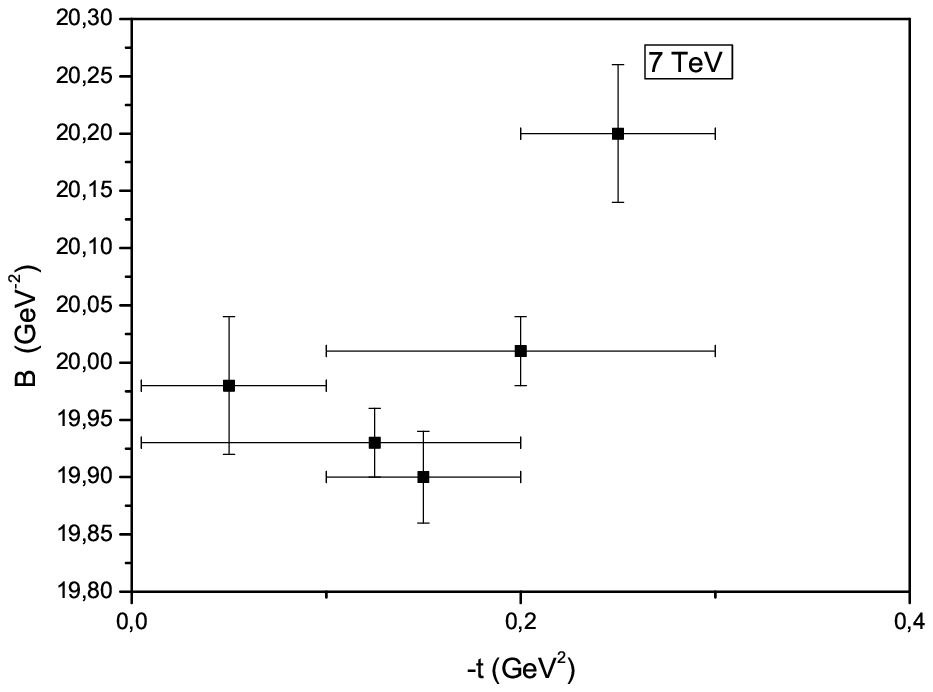} }
\caption{Local slopes $B(t)$ calculated at 7 TeV up to $\left|t\right|=0.3$ GeV$^2$} (see also Table 6. in \cite{TOTEM1}). 
\label{fig:B7TeV}
\end{figure}

 An immediate conclusion is that in the calculation of $B(t)$ the result depends on the bins in $t$ chosen. Generally speaking, the bins can be chosen arbitrarily: small (containing at least three data points) or large. They may be chosen in a touching sequence or overlap. The latter option (so-called overlapping bins method (OBM)) was studied in details in a number of papers \cite{JLL, Desgrolard, KL95, Kontros, CudellPR}, whose ideas and results are quoted in the Appendix.     

 Below we discuss in more details all these results and make predictions for $8$ TeV. A preliminary version of this study was presented \cite{Protvino} at the Protvino Conference on High-Energy Physics in June, 2014.

To start with, we recalculate the local slope 
with account for both  the statistical and systematic errors. To this end we will choose the compiled data from \cite{data}.
 At the LHC we include only data from the first cone \cite{TOTEM1,Mirko} for $\left|t\right|<0.2 GeV^2$.

For the ISR data we have chosen the Amaldi {\it et al.} data \cite{Amaldi} for center of mass energies  23.5, 30.8 and 44.7 GeV. The results are quoted in Table~\ref{TabIII}.

\begin{table}
\centering
\caption {Local slopes  B(t) recalculated  with different exponentials for the ISR  data  {\cite {Amaldi}}.}
 \label{TabIII}
\begin{tabular}{|c|c|c|c|c|c|}  \hline 
  E c.m.,    & $\left|t\right|$-range, & B,           & err.&  $\Delta$B,  & err.           \\
   GeV       & $(\mathrm{GeV}^2)$               & $(\mathrm{GeV}^{-2})$ &     &  $(\mathrm{GeV}^{-2})$  & $(\mathrm{GeV})^{-2}$    \\ \hline
    23.5     & 0.05 - 0.102            & 11.5         & 0.7 &               &                \\
             & 0.138 - 0.238           & 10.2         & 0.4 & 1.22          &   1.1          \\ \hline    
      32.0   & 0.05 -  0.094           & 11.6         & 0.4 &               &                \\
             & 0.138 - 0.240           & 10.9         & 0.3 & 0.7           &   0.7          \\ \hline
      44.7   & 0.05 - 0.096            & 13.3         & 0.3 &               &                \\
             & 0.138 - 0.238           & 10.6         & 0.2 &  2.7          &   0.5          \\ \hline
 \end{tabular}
\end{table}

The results of the calculations  coincide within errors with the data from Table~\ref{TabI} of 
Ref.~ \cite{BB72}.

 There is a gap between adjacent bins in the ISR data, seemingly increasing the  `break' of the slope (see Fig. \ref{fig:ISRslope}). 
For example, the choice of the bins adopted in Ref. \cite{BB72} is not unique. For example, by selecting the location of the second bin at CME= 23.5 GeV as $  0.5<\left|t\right|< 1.0$, the averaged slope within this bin will be $B = 9.5 \pm 0.1$, implying $\Delta B = 1.9 \pm 0.8$, which is more reliable than the results of Table~\ref{TabIII}. 
 
However, it is more natural to calculate the  `break'  for neighbouring bins in the vicinity of $\left|t\right| \approx 0.1$, close to the "break" we are scrutinizing. 
The local slopes calculated for the ISR and SPS  data \cite{BozzoUA4,Amaldi} (compiled in \cite{data}) for bin intervals being near the same as in \cite{BB72} are shown in Table~\ref{TabIII}.

It is interesting to study whether this effect (the `break' of the local slope) persists up to the LHC energies. To this end we construct a similar plot for the slopes  $B(t)$ (and compare it with Table 6. of \cite{TOTEM1}) for the differential cross sections measured at  $\sqrt{s}$ = 7 TeV in the interval $0.005 <\left|t\right| < 0.3$ (Fig. \ref{fig:B7TeV}).
 
For clarity sake we performed the calculations of local slopes in adjacent bins around $\left|t\right|=0.1 GeV ^2$ for nearly  the same length as in \cite{BB72}: the first bin is $~0.05 <\left|t\right| <~0.1$; the second one is $~0.1 <\left|t\right|<~0.14.$ As a result, the value of the `break' $\Delta B$ varies within $0.5 - 1.2$ GeV$^{-2}$ for the ISR and TOTEM energies (see Table~\ref{TabIV}).
\begin{table}
\centering
\caption{Calculated break $\Delta B$ from adjacent bins for the ISR and TOTEM data. }
 \label{TabIV}
\begin{tabular}{|c|c|c|c|c|c|}  \hline
  E c.m., GeV &   $t$-bin,$\mathrm{GeV}^2$   & B, $\mathrm{GeV}^{-2}$    & err.  &  $\Delta$B, $(\mathrm{GeV}^{-2})$ & err. \\ \hline
              & 0.05-0.094  & 11.5 &  0.7 &     &      \\
    23.5      & 0.094-0.138 & 10.2 &  0.9 & 1.3 & 1.6  \\ \hline
              & 0.05-0.094  & 11.6 &  0.4 &     &      \\
    30.7      & 0.090-0.138 & 11.2 &  0.6 & 0.4 & 1.0  \\ \hline
              & 0.05- 0.094 & 13.1 &  0.3 &     &      \\
    44.7      & 0.094-0.136 & 12.3 &  0.5 & 0.8 & 0.8  \\ \hline
              & 0.05- 0.10  & 15.3 &  1.2 &     &      \\
    546.      & 0.10-0.138  & 13.9 &  1.5 & 1.4 & 2.7  \\ \hline
              & 0.046-0.091 & 19.8 &  0.7 &     &      \\
    7000      & 0.091-0.137 & 19.3 &  0.9 & 0.5 &  1.6 \\ \hline
              & 0.05-0.095  & 19.8 &  0.2 &     &      \\
    8000      & 0.095-0.137 & 18.8 &  0.4 & 1.0 &  0.6 \\ \hline \hline
              & 0.05-0.095  & 19.8 &  0.2 &     &      \\
    8000      & 0.095-0.189 & 19.1 &  0.3 & 0.7 &  0.5 \\ \hline
   \end{tabular}
\end{table}

This estimate of the "`break"'  is less reliable, but shows the trend with energy. One concludes from Table~\ref{TabIV} that the break does not diminish with energy.

\section{Physics of the "break" phenomenon}

\begin{figure}[ht]
\center{\includegraphics[width=.9\textwidth]{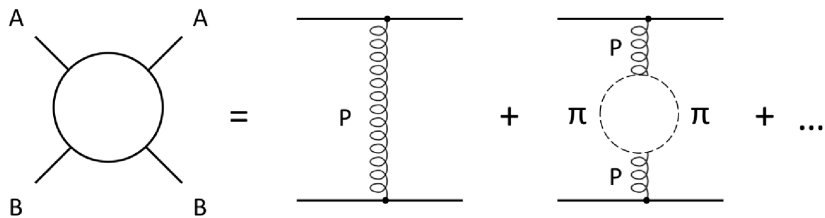} }
\caption{Feynmann diagram for elastic scattering with a $t$-channel exchange containing a branch point at $t=4m_{\pi}^2$}. 
\label{fig:diagram}
\end{figure}

The physics of the phenomenon was explained in Ref. \cite{Cohen}. The "break" (in fact a smooth concave over the linear exponential, approximated by two linear exponentials (cf. \cite{Mirko} ). This structure is due to the lowest two-pion exchange in the $t-$ channel required by $t-$channel unitarity \cite{Collins}, see Fig. \ref{fig:diagram}. The threshold appears at 
$t=4m_{\pi}^2\approx 0.08$ GeV$^2$, which is the mirror (with opposite sign of $t$) position of the "break" of the cone. We recall that any analytic function (here, the scattering amplitude) 
is sensitive to "mirror reflection" of its singularities (here, the $4m_{\pi}$ branch point in the amplitude). 

According to the ideas of duality  \cite{Collins, Fortschritte}, the singularities enter the amplitude through Regge trajectories. Below, following Ref. \cite{Cohen}, we present a model amplitude  realizing this principle and reproducing the observed "break".     

The $t-$ channel threshold shown in Fig. \ref{fig:diagram} may enter both through leading (Pomeron, Odderon) or non-leading ($f, \omega$) trajectories. 
While at the LHC, the low-$|t|$ are dominated completely by the Pomeron contribution (whatever it be!) \cite{JLL}, at the ISR energies, secondary Reggeons are not negligible, at least in  nearly forward scattering. 

A cut Pomeron trajectory including the lowest-lying $2m_{\pi}$ cut may be 
approximated by \cite{Cohen, Fortschritte}: 

\begin{equation}
\alpha(t) = 1+\delta +\alpha't - \gamma\left(\sqrt{4m_{\pi}^2 -t}-2 m_{\pi}\right),
\label{eq:tr2}
\end{equation}

In the next Section we extrapolate in $s$ the forward cone from the ISR to the LHC energies.
This is not a trivial task since a detailed fit requires the inclusion at "low", ISR energies the contribution from at least four trajectories, namely that of the Pomeron, evetntually the Odderon, and two secondary Reggeons, $f$ and $\omega$. Postponing this discussion to a forthcoming detailed analyses, here we use a single "effective" trajectory that at the low-energy part mimics all contributions mentioned (at LHC energies it is the Pomeron alone, see Ref. \cite{JLL}). 

\section{Extrapolating from the ISR to the LHC}

To extrapolate the cross-section (or just the slope) from "low" (ISR) energies to those at the LHC, we use a simple single Regge pole amplitude with an "effective" trajectory that is close to the Pomeron dominating the high-energy region. The intercept of the trajectory $\alpha(0)=1+\delta$, following the Donnachie and Landshoff approach \cite{DL} to high-energy phenomenology, will be set slightly above $1$.The relevant scattering amplitude reads

\begin{equation}\label{GP}
A(s,t)=g s\tilde{}^{-\Delta} e^{bt} s\tilde{}^{\alpha(t)},\ \ s\tilde{}= -i\frac{s}{s_0},
\end{equation}

\begin{equation}\label{norm}
\frac{d\sigma}{dt}|_{th}={\pi\over s^2}|A(s,t)|^2 .
\end{equation}

We use a representative Pomeron trajectory, namely that with a two-pion square-root threshold, Eq.~(\ref{eq:tr2}), required by $t-$channel unitarity and accounting for the small-$t$ `break' \cite{Cohen}.

The normalized `experimental' points of $R(t)$ are defined as:  
 \begin{equation}
R(t) = ((d\sigma/dt)_{exp} - (d\sigma/dt)_{lin}) /(d\sigma/dt)_{lin},
\label{eq:R1}
\end{equation}
where
\begin{equation}\label{eq:extrapolate}
\biggl(\frac{d\sigma}{dt}\biggr)_{lin}=ae^{bt}
\end{equation}   

The theoretical values of $R(t)$ are calculated from 
\begin{equation}
R(t) = ((d\sigma/dt)_{exp} - (d\sigma/dt)_{th} )/(d\sigma/dt)_{th}.
\label{eq:R2}
\end{equation}

\begin{figure}[ht]
\center{
 \includegraphics[trim = 8mm 3mm 15mm 5mm,clip,width=.32\textwidth]{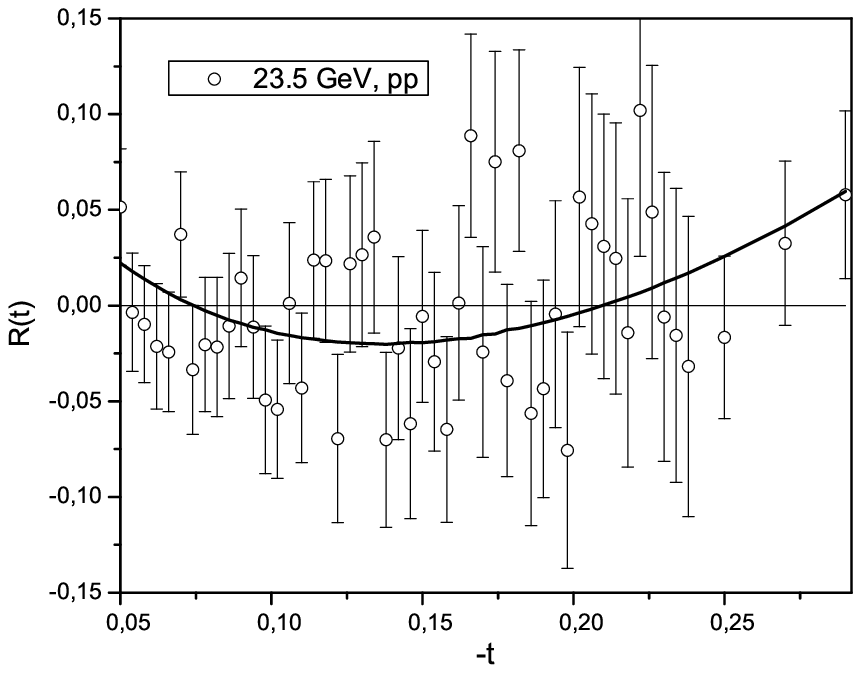}
 \includegraphics[trim = 8mm 3mm 15mm 5mm,clip,width=.32\textwidth]{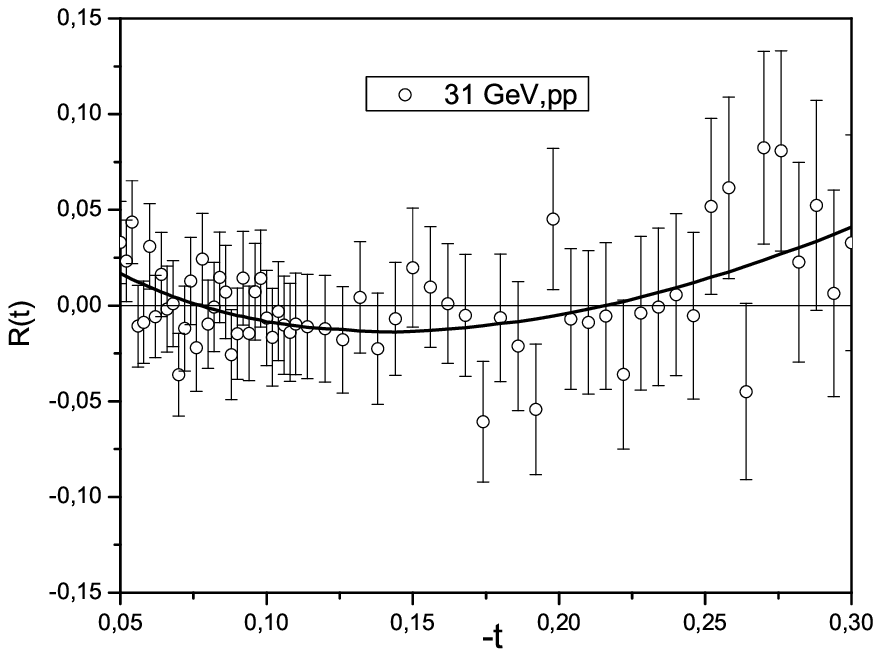}
 \includegraphics[trim = 8mm 3mm 15mm 5mm,clip,width=.32\textwidth]{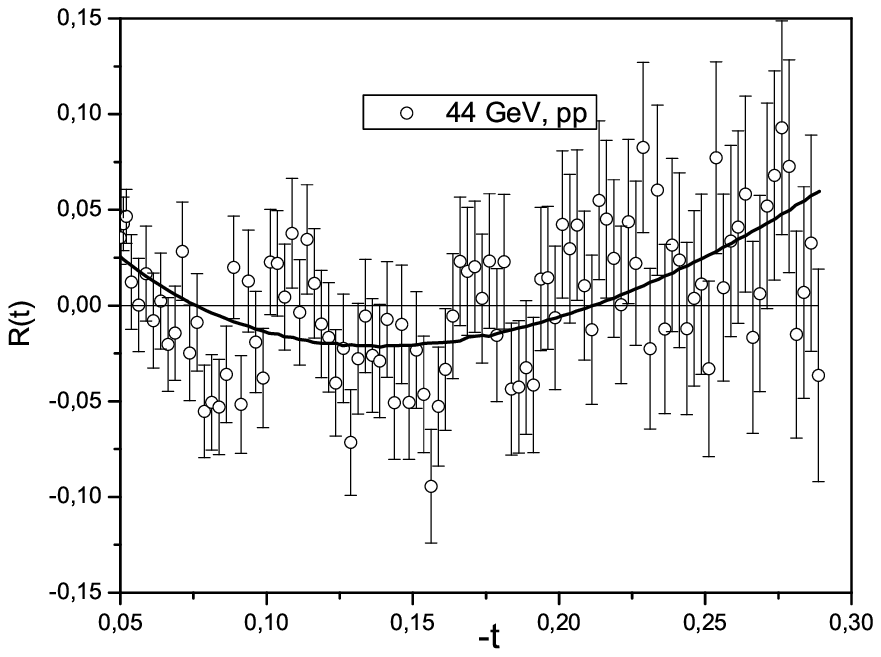}
}
\caption{ $R(t)$ calculated at ISR  energies} 
\label{fig:ISR_R}
\end{figure}

Those for $(d\sigma/dt)_{th}$ correspond to the solid curve in Fig. \ref{fig:ISR_R} calculated as the best fit to the experimental differential cross sections  for a given energy with the free parameters  $g_p$, $b_p$, $\alpha_2$  and fixed  $\delta=0.08$ and $\alpha_1=0.23$; $t_0=4m^{2}_{\pi}$, where
$m_{\pi}$  is the pion mass. 

For all energies, the value of $R(t)$ clearly demonstrates concavity at -t = 0.1 $GeV^2$, which is in qualitative agreement with the "experimental" one in $R(t)_{exp}$.

\begin{figure}[ht]
\center\includegraphics[width=.4\textwidth]{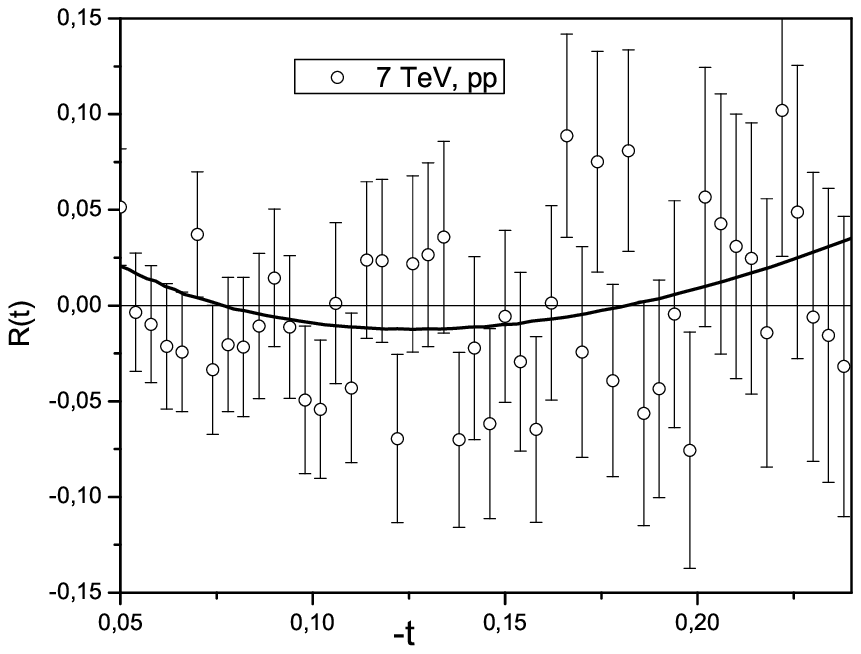}
\includegraphics[width=.4\textwidth]{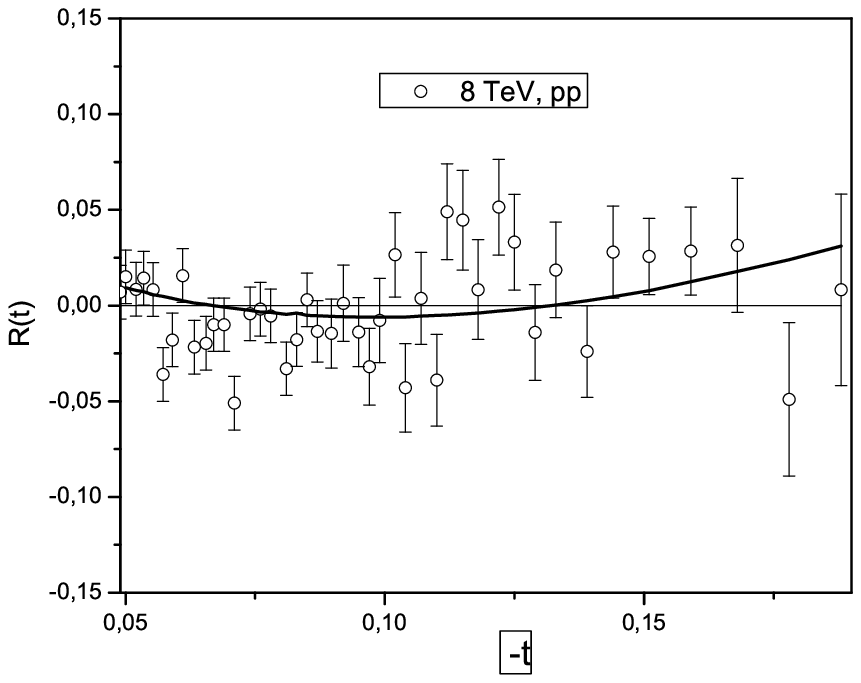}
\caption{ $R(t)$ calculated at 7 and 8	TeV.} 
\label{fig:TOTEM_R}
\end{figure}

The value $R(t)\neq  0 $ around  $-t= 0.1$ GeV$^2$ means that the experimental data $(d\sigma/dt)_{exp}$ are not compatible with a simple exponential.

\section{Tiny oscillations?}
Besides the "break" discussed above, small-$|t|$ oscillations on the smooth exponential cone may also be present. They were discussed in a number of papers - theoretical and experimental \cite{ Kontros}, \cite{BH92,BH94,Selyugin,Denisov,Ezhela}. Since the amplitude of the possible oscillations appear to be close to the error bars, it is still not clear whether this is an experimental fact or an artefact. In Refs.~\cite{KL95,Kontros} the low-$t$ data were fitted
to a model which, apart from the $t_0=4m_{\pi}^2$ cut, contains also an oscillating term (in the cross section or in the slope $B(t)$):
\begin{equation}
B(t)=\alpha'\frac{\gamma}{2\sqrt{t_0}-t}+a\cos (\omega t+\phi),
\end{equation}  
where $a,\ \omega$ and $\phi$ are fitted parameters. The result is shown in Figs. 8 and 9. The dashed curve are calculated from Eqs. (1)-(6) and they correspond to the smoothed part of (7), see
Ref. \cite{Kontros}. Note that, since the derivative of an oscillating function is also oscillating, it makes little difference whether one is fitting the cross sections or the slope.

The overlapping bins method (see Appendix) may be extremely usefu in performing this delicate analysis. Since the earlier (theoretical and experimental) results are still inconclusive, new measurements (e.g. those by the Denisov group in Protvino \cite{Denisov}) are very important to shed new light on this phenomenon.
   
The physics behind the possible low-$t$, small amplitude oscillations may be related to those at large impact parameters. As discussed in Ref. \cite{Kuraev}, large-distance residual Van der Waals forces may be responsible for these oscillations.

\begin{figure}[ht]
\center 
\includegraphics[trim = 11mm 3mm 15mm 5mm,clip,width=.32\textwidth]{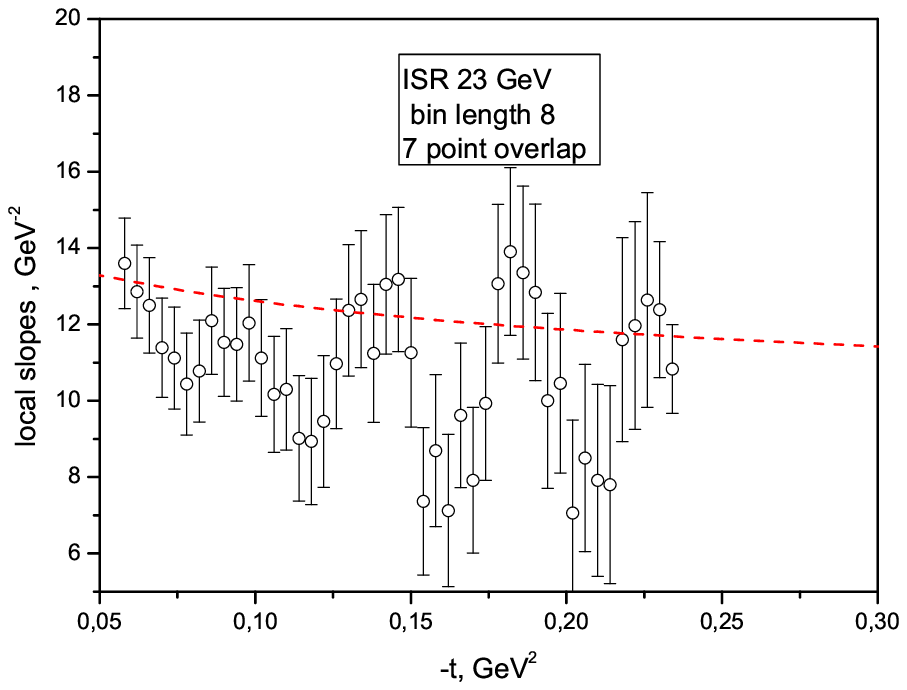}
\includegraphics[trim = 11mm 3mm 15mm 5mm,clip,width=.32\textwidth]{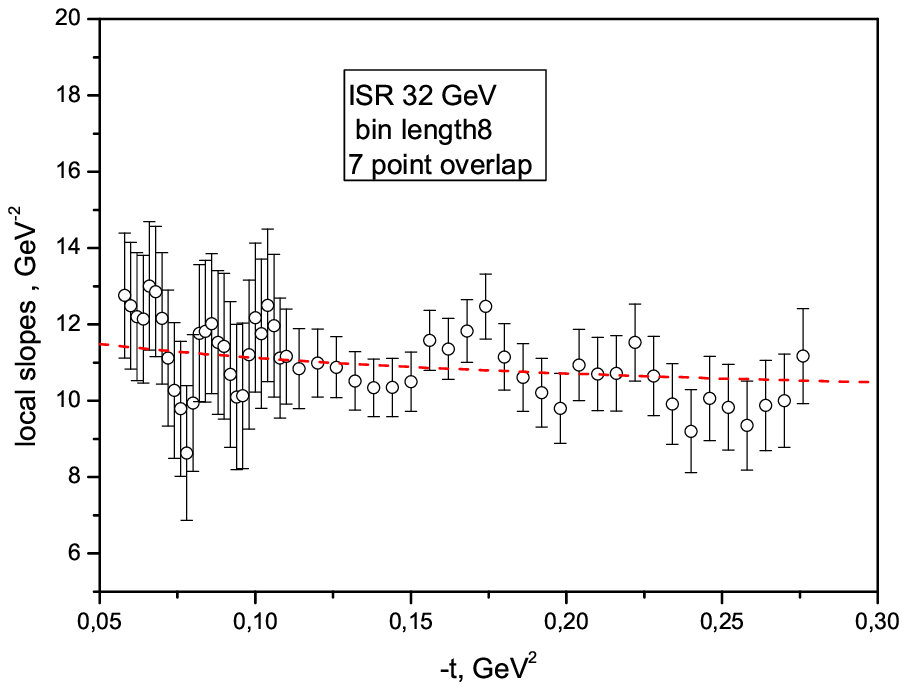}
\includegraphics[trim = 11mm 3mm 15mm 5mm,clip,width=.32\textwidth]{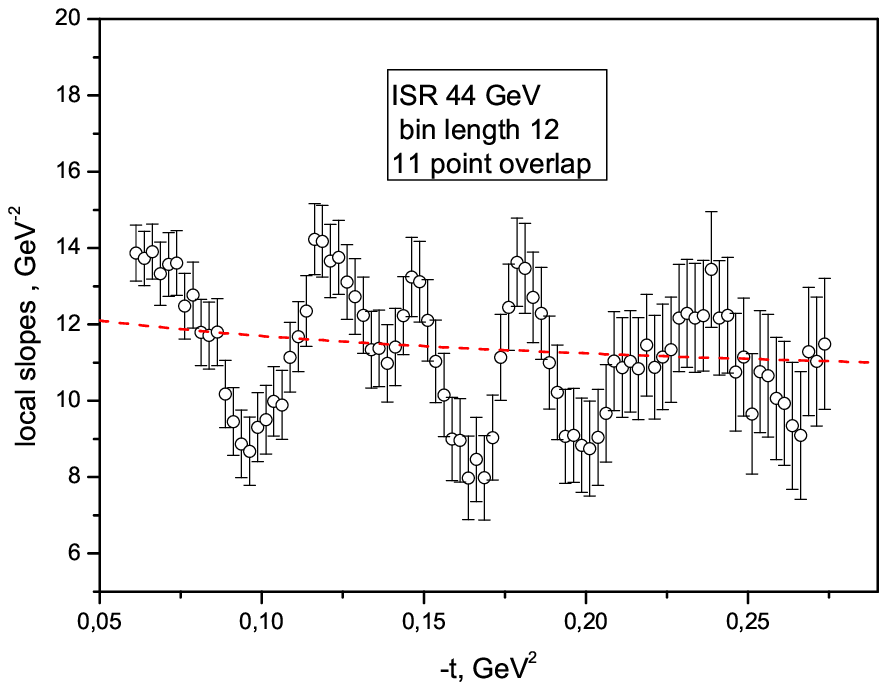}
\caption{Local slope $B(t)$ calculated at $23.5$ GeV, $32.5$ GeV and 45 GeV for  overlapping bins.}
 \label{fig:B23}
\end{figure}

\begin{figure}[ht]
\center\includegraphics[width=.4\textwidth]{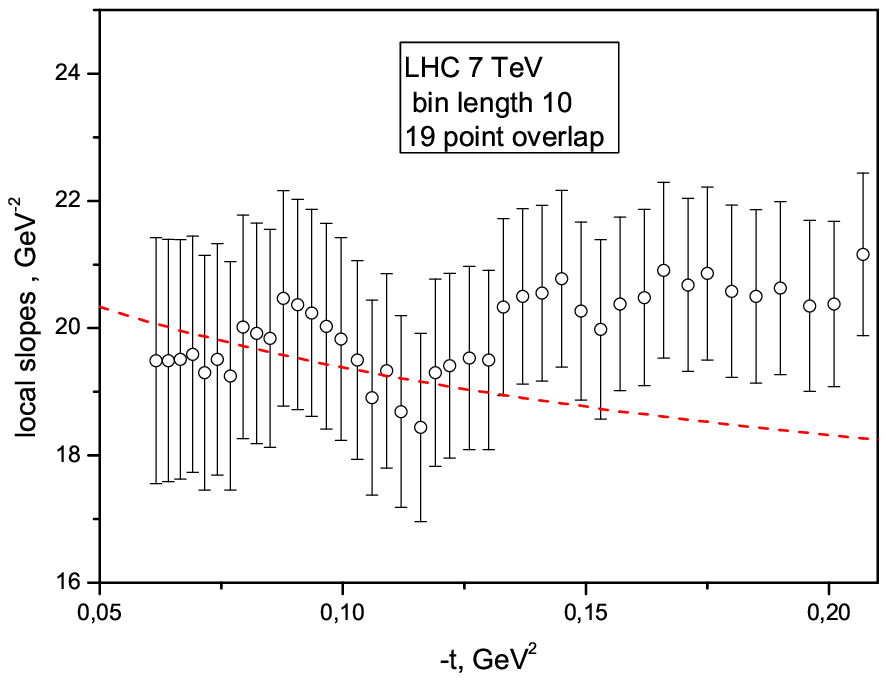}
\includegraphics[width=.4\textwidth]{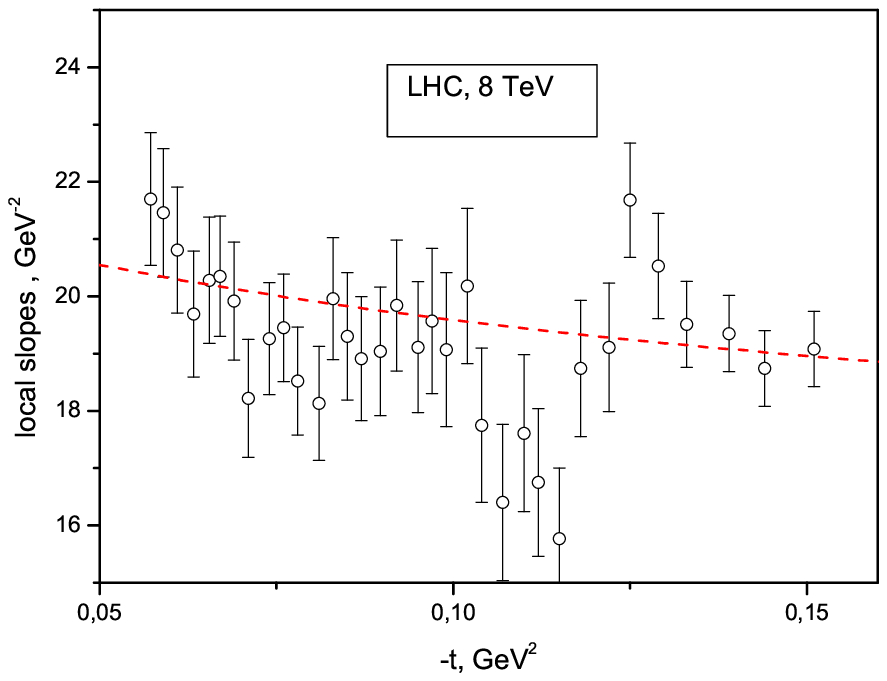}
\caption{Local slopes with  overlapping bins for the data at $7$ TeV and 8 TeV. }
\label{fig:7overlap}
\end{figure}

\section{Conclusions}
We conclude that the "break" observed \cite{Mirko, Kaspar, Csorgo} by TOTEM near $-t=0.1$ GeV$^2$ at $8$ GeV is a "recurrence" of a similar structure seen in 1972 at the ISR. 

Note that the break was not seen at the Tevatron at $1.8$ TeV. Possible reasons for the
non-appearance of the break in $\bar pp$ may be relaled to the Odderon contribution masking it.

While the change of the slope $B(t)$ near $-t\approx 0.1$ appears to be a universal and well established phenomenon (although its energy (in)dependence needs better understanding), the status of the tiny oscillations is still ambiguous. It may be that the "break" near $-t=0.1$ GeV$^2$ is part of the oscillations \cite{Kontros}.

\section*{Acknowledgements}
Useful discussions and correspondence with Mirko Berretti, Tam\'as Cs\"org\H o, Simone Giani and Jan Kaspar are acknowledged. We thank Andrii Salii for his help in preparing the manuscript.      
       
 \section{APPENDIX: The overlapping bins method (OBM)}\label{Sec:OB} 
The fine structure of the diffraction peak in the
differential $\overline{p}p-$ and $pp-$ elastic cross section was first observed  in  \cite{White} at the ISR \cite{BB72}, followed by the UA4/2 experiment \cite{BozzoUA4} by normalizing the differential cross-section to the
smoothly varying background in the impact parameter
representation  \cite{BH94}. In \cite {BH92} an attempt was made to relate the
observed structure near $|t|=0.1$ $GeV^2$ to the variation of the
opacity in $b-$space, probably reflecting the density oscillation
in matter. The possible existence of oscillations with even
smaller periods was discussed by several authors
\cite{PG97}.

In ref. \cite{KL95} an entirely different method of
identifying the fine structure  in the $\overline{p}p-$ and $pp-$ elastic
scattering was proposed. The method, unlike the that of \cite{Schiz}, is based on the use of overlapping bins
of local slopes. Small oscillations, over the exponential done, with a charactiristic period
were discovered. It is obvious that, in order to determine the
nature and periods of the oscillations, one first has to
improve the reliability of the initial information contained in
the experimental data by suppressing the influence of statistical
fluctuations. This problem can be settled by means of the well
known method of maximum entropy \cite{Max91}, used in many areas
of physics. Recently it has been applied \cite{DKL} to the hadron
scattering data.

The method is based on the use of overlapping bins
of local slopes.

To check the expected behaviour of the slope
\begin{equation}
B(s,t)=\frac{d}{dt}\ln \left( \frac{d\sigma(s,t)}{dt}\right)
\label{eq10}
\end{equation}
over $t$ one must operate with its "experimental" value.

Provided that
\begin{equation}
\left( \frac{d\sigma}{dt}\right)_i=\left|a_ie^{b_it} \right|^{2}
\label{eq11}
\end{equation}
has been measured for a given $s$ at $N$ $|t|$-points lying in
some interval $[|t|_{min},|t|_{max}]$, we adopt the following
procedure. First, we divide this interval into subintervals or
elementary "bins" (with $n_{b}$ measurements in each of them,
assumed for simplicity to be the same for all bins). Once the
first bin is chosen, the second bin is obtained from the first one
by shifting only one point of measurement (of course, one could
shift it by any number of points less or equal to $n_{b}$, the
shift of one point is the minimal one giving rise to the maximal
number of overlapping bins). The third bin is obtained from the
second bin by shifting of one data point etc. Thus, we define
$N-n_{b}+1$ overlapping bins for a given $s$. For each ($k$-th)
bin, $n_{b}$ must be large enough and its width (in $|t|$) -
small enough to allow fitting $\left(\frac{d\sigma}{dt}\right)$
with the simplest form directly involving the $t$-slope $b$
(\ref{eq11}).The parameter ${b}$ represents the value of the $t$-slope
$B\left(<t>_{k},s\right)$ "measured" at $s$ and "weighted
average" $<t>_{k}$. This yields the "experimental" values of $b_{k}(s,t_{k})$ with the corresponding
standard errors determined in the fit of (\ref{eq11}) to the
data. Then the procedure is to be repeated for all bins and
ultimately for the other $t$'s at which the $\left(\frac{d\sigma}{dt}\right)$ have been measured. 
A regular structure in the local slope of diffraction cone
$B(s,t)$ was found by procedure of the overlapping bins described
above and applied to experimental data \cite{KL95}. However if one has the
bins of $\sim 10$ points and shifts them at each step by one
only, the overlap my be so strong that the information from the
neighbors is extremely correlated and one can ascribe the regular
tendency of the final plot increase (or decrease) in many
neighboring bins to thus correlation.

To resolve these doubts, the local slopes were re-evaluated
with the help of the Overlapping Bins Metod (OBM) by the LSQ method with so called
"non-independent $y$'s" \cite {PartPhys}.

Rather than minimizing the functional

\begin{equation}
\label{eq17}s=\sum_1^N\left( \frac{f\left( t_i,{\bf a}\right)
-y_i}{\Delta y_i} \right) ^2
\end{equation}
the form 

\begin{equation}
\label{eq18}s=\sum_{ij}^N\left( f\left( t_i,{\bf a}\right)
-y_i\right) w_{ij}\left( f\left( t_j,{\bf a}\right) -y_j\right)
\end{equation}
can be used.

In the framework of the OBM calculus

\begin{equation}
\label{eq19}s=\sum_{j=1}^{N}\sum_{i=j}^{j+n-1} \frac{\left(
f\left( t_i,a_j\right) -y_i\right) ^2}{\Delta
y_i^2}-\sum_{j=1}^{N}\sum_{i=j+1}^{j+n-2} \frac{\left( f\left(
t_i,a_j\right) -y_i\right) \left( f\left( t_i,a_{j+1}\right)
-y_i\right)}{\Delta y_i^2}.
\end{equation}
Calculations using correlation
over {Eq.~(\ref{eq19})} for the same set of points indicate that the
errors $\Delta b_i$ are reduced and oscillations are revealed more
distinctly, i.e. the relation between the value of errors and the
amplitude of oscillation can be improved. 




\begin{thebibliography}{99}

\bibitem{TOTEM2} TOTEM Collab., G. Antchev et al., arXiv: 1110.1395.



\bibitem{TOTEM1} TOTEM Collab., G. Antchev {\it et al.}, Europhys. Lett. {\bf 101} (2013) 21002.  

\bibitem{Mirko} Mirko Berretti, {\it TOTEM results}, presented at "Diffracion 2014", Primosten, Sept. 2014,   
to be publ. in the Conference Proceedings. 


\bibitem{Simone} Simone Giani, Contribution at the Xth Workshop on Particle Correlations and Femtoscopy (WPCF-2014), Gyongyos, Hungary, Sept. 2014.

\bibitem{Kaspar} Jan Ka\v{s}par,  {\it TOTEM results on elastic scattering and total cross sections}, presented at the International Workshop on High-Energy Physics, Protvino, June 2014:  
to be published in the Conference Proceedings.
	

\bibitem{Csorgo} Fabrizio Ferro, {\it TOTEM elastic scattering etc.}, Contribution at the "Low-x" Meeting in Kyoto, 2014.

\bibitem{Protvino} L. Jenkovszky, presentation at the HEPFT2014 conference:\\
{ https://indico.cern.ch/event/269671/other-view?view=cdsagenda},\\
to be published in the Proceeings.


\bibitem{BB72}
B. Barbiellini et al, \textit{Phys. Lett.} \textbf{B 39} (1972)
663.

\bibitem {BozzoUA4} M.Bozzo at al, \textit{Phys. Lett.} \textbf{B 147} (1984) 385


\bibitem {Desgrolard} P. Desgrolard, J. Kontros, A.I. Lengyel, E.S. Martynov. Nuovo Cim. {\bf 110A}, 615 (1997).


\bibitem{JLL}	L.L. Jenkovszky, A.I. Lengyel, D.I. Lontkovskyi, {\it The Pomeron and Odderon in elastic, inelastic and total cross sections at the LHC}, Int. J. Mod. Phys. A26 (2011) 4755-477; arXiv:1105.1202. 


\bibitem{Cohen} G. Cohen-Tannoudji {\it et al.}, Lettere Nuovo Cim. {\bf 5} (1972) 957.









\bibitem{KL95}J. Kontros, A. Lengyel, in {\it Strong Interaction at Long Distances},  L.L.~Jenkovszky, Ed. (Hadronic Press, Palm Arbor) (1995) 67.

\bibitem{Kontros} J.E.~Kontros, A.I.~Lengyel, Ukr. J. Phys., {\bf 40} 263 (1995); {\it ibid.} {\bf 41} 290 (1996).


\bibitem {CudellPR} J.R. Cudell, A. Lengyel, E. Martynov. Phys.Rev.D {\bf 73} (2006) 034008.
\bibitem{data} File with the data "alldata-2.zip" is available at the address:\\ http:/www.thwo.phys.ulg.ac.be/alldata-v2.zip.
\bibitem{Amaldi}  R. Amaldi et al. Nucl. Phys. {\bf B166} (1979) 301.




\bibitem{Collins} P.D.B. Collins, {\it An introduction to Regge pole \& high energy physics}, Cambridge University Press, 1977. 








\bibitem{Fortschritte} G. Cohen-Tannoudji {\it et al.}, Fortschritte der Physik, {\bf 21} (1972) 427.

\bibitem{DL} A. Donnachie, P.V. Landshoff. Nucl. Phys. B {\bf 267}, 690 (1986). 


\bibitem{White}
J.N. White, \textit{Nucl. Phys.} \textbf{B 51} (1973) 23.



\bibitem{BH92}S. Saul Barshay, Patrick Heiliger, Dieter Rein. Z.Phys.C. 56, 77 (1992); 
A. Arnold, S. Barshay, \textit{Nuovo Cim.} \textbf{35A} (1976)457;

\bibitem{BH94}S. Barshay, P. Heiliger, \textit{Z. Physik} \textbf{C 64} (1994)675 and references herein.

\bibitem{Selyugin} O.V. Selyugin, {\it Oscillation of the hadronic amplitude at small transfer momenta} in the Proceedings of the conference "Hadrons-95", edited by G. Bugrij et al., Kiev, 1994, p. 65; Phys.Lett. B333, 345 (1994). 

\bibitem{Denisov} Yu.M. Antipov  {\it et al.}, $\pi^{+}p -. K^{+}p$ - and $pp$ Elastic Scattering in the Momentum Range 29-65 GeV/ñ, Serpukhov, 1976, p.44. (IHEP 76-95 preprint). 



 \bibitem{Ezhela} V. Ezhela et al. Diffraction 2002: Proceedings of the NATO Advanced Research Workshop, Edited by R. Fiore {\it et al.}, Lluwer Academic Publishers, v.101, 2002, p. 47.



\bibitem{BSW93}
C. Bourrely, J. Soffer, T.T. Wu, \textit{Phys. Lett.} \textbf{B 313} (1993) 195.
\bibitem{CA93}
C. Augier et al, \emph{UA4/2 Collaboration}, \textit{Phys. Lett.}
\textbf{B 316} (1993) 448.
30
\bibitem{PG97}
P. Gauron et al, \textit{Phys. Lett.} \textbf{B 397} (1997) 305.
21



\bibitem{Schiz} A. Schiz et al. Phys. Rev. {\bf 172}, 1413 (1968).

\bibitem{CERN}
CERN Computer Center Program Library, {\bf D506}.

\bibitem{PartPhys} Review of Particle Physics, Phys. Rev. D, {\bf 54} (1996) 161.



\bibitem{Kuraev}  E.A. Kuraev, P. Ferro and L. Trentadue, {\it Possible manifestation of long range forces in high-energy hadron collision}, JINR Preprint E3-97-95 (Dubna) 1997.

\bibitem{Max91}
B. Buck and V.A. Macaulay (eds), {\it Maximum Entropy in Action}, (Oxford: Clarendon) (1991).


\bibitem{DKL}
O. Dumbrajs, J. Kontros, A. Lengyel, \textit{J. Phys.} \textbf{G26} {2000}{1321}.




\end{thebibliography}
\end{document}